\documentstyle[aaspp4,epsfig,psfig]{article}

\def\la{\mathrel{\mathchoice {\vcenter{\offinterlineskip\halign{\hfil
$\displaystyle##$\hfil\cr<\cr\sim\cr}}}   
{\vcenter{\offinterlineskip\halign{\hfil$\textstyle##$\hfil\cr
<\cr\sim\cr}}}
{\vcenter{\offinterlineskip\halign{\hfil$\scriptstyle##$\hfil\cr
<\cr\sim\cr}}}
{\vcenter{\offinterlineskip\halign{\hfil$\scriptscriptstyle##$\hfil\cr
<\cr\sim\cr}}}}}

\def \ml{ M_{ac} } 
\def \pmin{ P_{min}^{obs} }
\def \msun{M_{\odot}}
\begin{document}
 
\title{Neutron stars with submillisecond periods: a population of high
mass objects?}
                                  
\author{Luciano Burderi\altaffilmark{1}, Andrea Possenti\altaffilmark{2},
Monica Colpi\altaffilmark{3}, Tiziana Di Salvo\altaffilmark{1},
Nichi D'Amico\altaffilmark{4}}

\altaffiltext{1}{Dipartimento di Scienze Fisiche ed Astronomiche 
dell'Universita', via Archirafi 36, 90123 Palermo, Italy}
\altaffiltext{2}{Dipartimento di Astronomia dell'Universita', via Zamboni 33,
40126 Bologna, Italy}
\altaffiltext{3}{Dipartimento di Fisica dell'Universita', via Celoria 16,
20133 Milano, Italy}
\altaffiltext{4}{Osservatorio Astronomico di Bologna, via Zamboni 33,
40126 Bologna, Italy}

\vskip 25pt
\centerline{ \it To appear in The Astrophysical Journal, June 1999}
\vskip 25pt
\begin{abstract}

Fast spinning neutron stars, recycled in low mass binaries, 
may have accreted a substantial amount of mass.  
The available relativistic measurements of neutron star masses,
all clustering around $1.4\, M_{\odot},$ however refer mostly 
to slowly rotating neutron stars 
which accreted a tiny amount of mass during  evolution in a massive  
binary system. 

We develop a semi--analytical model for studying the evolution of 
the spin period $P$ of a magnetic neutron star as a function of the baryonic
mass load $M_{ac}$; evolution is followed down to submillisecond periods 
and the magnetic field is allowed to decay significantly before the end 
of recycling. We use different equations of state and include rotational 
deformation effects, the presence of a strong gravitational field and of 
a magnetosphere. For the non--magnetic case, comparison with numerical 
relativistic codes shows the accuracy of our description.

The {\it minimum} accreted mass requested to spin--up a {\it magnetized}  
$1.35~M_\odot$--neutron star at a few millisecond is $\sim 0.05 M_\odot$,
while this value doubles for an unmagnetized neutron star. 
Below 1 millisecond the request is of at least $\sim 0.25 M_\odot.$
Only highly non--conservative scenarios for the binary evolution could prevent 
the transfer of such a mass to the compact object. Unless a physical mechanism
limits the rotational period, there may exist a yet undetected population 
of massive submillisecond neutron stars. 
The discovery of a submillisecond neutron star would imply a lower limit
for its mass of about $1.7~M_\odot$.
\end{abstract}


\section{Introduction}
The evolution of a neutron star (NS) in a binary system 
differs considerably  in the case the donor star is a high 
mass giant star in a HMXB 
or is a lighter star in a LMXB (see {\it e.g.} Verbunt 1993; Lipunov 1992).
It is only in the former case that the formation of a  
pulsar (slowly rotating at $P>50$ ms) in a relativistic 
NS-NS binary allows for the 
accurate estimate of the masses of the two components (Portegies, 
Zwart \& Yungelson 1998). Collecting all data for the known binary pulsars,
Thorsett \& Chakrabarty (1998) found an average mass of $1.35~M_\odot$ 
(with a very narrow spread $\sigma = 0.04~M_\odot$) for 
both NSs in these systems. These values are fully compatible with the two
hypotheses: that  NSs at birth possess a canonical mass of about 
$1.4~M_\odot$ (Woosley \& Weaver 1986; but see Timmes, Woosley \& Weaver 1996)
and that the HMXB phase is shortlived so that at most $10^{-2}~M_\odot$ 
can be accreted onto the oldest nonpulsating NS (Bhattacharya \& 
van den Heuvel 1991). On the other hand, it is widely accepted (Phinney \& 
Kulkarni 1994, van den Heuvel \& Rappaport 1992) that the formation of 
millisecond pulsars (MSPs; with period $P<10~$ms) takes place inside
LMXBs, where an old neutron star is spun--up by accretion torques. 
The mass load can be significant, in these systems, and, in principle, 
the NS can reach ultrashort periods $P \la 1$ ms (depending on the equation 
of state) on a time that is controlled by the evolution of the magnetic 
field (Possenti {\it et al.} 1998). Thus, there is a close relationship 
between the spin period $P,$ the gravitational mass and the magnetic 
field evolution.

Cook, Shapiro \& Teukolsky (1994a: CSTa) developed a relativistic code to
study the spin evolution of an unmagnetized NS accreting from 
the inner edge of a Keplerian disk. Here we reproduce their results 
using a semi--analytical model (described in section $\S~2$). Then we 
allow for the presence of a magnetosphere around the NS 
and explore the consequences on the 
relation between the baryonic mass load $M_{ac}$ and the spin rate $P$, 
considering two scenarios for the  magnetic field evolution. 
The minimum mass load requested to reach the submillisecond range is computed 
for a magnetic neutron star in section $\S~3$, adopting different equations
of state (EoSs). Observational tests are discussed  in section $\S~4$; 
in $\S~5$ we present our conclusions.

\section{Mass loading and spin evolution}
In this section we shortly describe the semi--analytical 
model used to compute the $P~$vs$~M_{ac}$ relation, which is estimated down 
to limiting periods $\la~1$ millisecond.
We first derive a simple equation describing the evolution 
of the gravitational mass $M_G$ as a function of the baryon load.
We then introduce  the  set of equations  for the evolution of 
the rotational frequency $\Omega$, including rotational effects on the NS
radius during recycling. The angular momentum transfer
rate is computed including the corrections due to a strong gravitational field.
The equations are solved numerically.

The code follows the evolution of an NS
of initial gravitational mass $M_{G,0} = 1.4 M_{\odot}$
which is accreting at a constant rate $\dot{m}_B$. 
The initial values of the NS baryonic mass $M_{B,0}$ and of
its radius $R_0$ are known once we specify the  EoS   (see
{\it e.g.} Cook, Shapiro \& Teukolsky 1994b; CSTb).
In our model, the gravitational and the baryonic masses $M_G$ and $M_B$ of 
the NS are related, at any given time, by  $M_G = M_B (1-\alpha/R),$
where $R$ is the NS radius circumferential radius ({\it i.e.} the proper
circumferential length at the NS equator divided by $2\pi$) and $\alpha$ is 
a constant evaluated using the initial set of parameters. Accordingly, 
\begin{equation}
M_{G} = M_{B} \left[1+\left(\frac{M_{G,0}}{M_{B,0}}-1\right)
\left(\frac{R_0}{R}\right)\right]
\end{equation} 
The decrease in radius due to the mass load is described using the simple 
relation $R \propto M_G^{-1/3}$ (for equilibria of degenerate nonrelativistic
neutrons): with this scaling  we estimate the non-rotating NS circumferential
radius $R$ at any stage of the accretion. The  evolution equation for  
$M_G$ thus reads: 
\begin{equation}
\dot{M}_G~=~\psi ~\dot{m}_B~=
~\left[ 1 + \left( \frac{M_{G,0}}{M_{B,0}} - 1 \right) 
\left( \frac{M_G}{M_{G,0}} \right)^{1/3} \right] \left[1-
\frac{M_B}{3 M_G} \left( \frac{M_{G,0}}{M_{B,0}} - 1 \right)
\left( \frac{M_G}{M_{G,0}} \right)^{1/3} \right]^{-1}
~\dot{m}_B.
\end{equation}

The mass increase is accompanied by the increase in the rotational
frequency $\Omega$, and at any time (i.e., at any current value of
$M_G$) we need to determine
whether  $\Omega$ is below or above the mass
shedding limit $\Omega_{max}$ that is the limiting angular frequency at
which the gravitational pull is balanced by the centrifugal forces. 
$\Omega_{max}$ is estimated using the classical expression 
\begin{equation}
\Omega_{max} = \Omega_K = 
\left( \frac{G M_G}{R_{\Omega,max}^3} \right)^{1/2} 
\end{equation}
where $R_{\Omega,max}$ denotes the circumferential radius
of the rotating NS which in turn depends upon $\Omega_{max}$.
$R_{\Omega,max}$ is estimated fitting the results of CSTb:
\begin {equation}
R_{\Omega,max}=1.5\,\,  R_0
\end{equation}
where $R_0$ is the radius of the static configuration at the current 
value $M_G$. In table 1 we compare equation (3) with the results of CSTb,
derived using a relativistic code for 13 different EoSs ($M_G = 1.4 M_\odot$).
The agreement is $\le 5\%$.  The maximum angular 
speed occurs when the NS equatorial speed reaches the  classical
Keplerian limit computed taking $M_G$ as gravitational mass and as radius
$R_{\Omega,max}$: the rapid rotation inflates and deforms the NS surface.
While the non-spherically-symmetric deformations (oblateness effects) are 
important in determining the radius of the spinning NS, they are almost 
irrelevant in changing the gravitational field felt by a test particle at 
the NS equator with respect to the corresponding spherical configuration.
This is probably due to the fact that most of the mass is concentrated well 
inside the external radius, where the deformation effects due to rotation are
almost negligible. This is predicted by the so called Roche Model  (see
{\it e.g.} Shapiro \& Teukolsky 1983), that, in the hypothesis outlined
above, conclude that the maximum expansion of 
a uniformly rotating star along its equator is a factor of 3/2, irrespective
of the EoS adopted.

If $\Omega<\Omega_{max}$, spin up can proceed and the circumferential
radius $R_{\Omega}$ of the rotating NS further increases.
$R_\Omega$ is computed using the relation
\begin{equation}
\log_{10}(R_\Omega/R_0) = {\frac{1-(1-\Omega/
\Omega_{max})^{1/3}}{6.75}}
\end{equation}
derived fitting the results of CSTb.
In Fig.~1 we report  the values  of $R_\Omega$ vs $\Omega$ from  
CSTb, for 5 EoSs ($M_B = 1.4 M_\odot)$ and the relation given by equation
(5). In the numerical scheme, $\Omega_{max}$ is firstly determined from (3) 
using equation (4) for $R_{\Omega,max}$.
Then (5) is used to calculate $R_\Omega$. 

\subsection{Unmagnetized case}
During the accretion process 
the specific angular momentum $l_{in}$ of the accreting matter is  
entirely transferred to the NS. From the angular momentum conservation
$d(I_{\Omega} \Omega)/d t = \dot{m}_B l_{in}$ we obtain
\begin{equation}
\dot{\Omega} = \frac{\dot{m}_B}{I_{\Omega}}~~
\left( l_{in} - \psi \Omega \frac{\partial I_{\Omega}}{\partial M_G} \right)
\left( 1 + \frac{\Omega}{I_{\Omega}} 
\frac{\partial I_{\Omega}}{\partial R_{\Omega}} 
\frac{\partial R_{\Omega}}{\partial \Omega} \right) ^{-1}
\end{equation}
where $I_{\Omega}$ is the moment of inertia of the neutron star. 
Taking into account the inflation of the radius caused by the rapid
rotation, we extrapolate a result of Ravenhall \& Pethick (1994)
obtained for static stars:
\begin{equation}
I_{\Omega} = 0.21 M_G R_{\Omega}^2 
\left( 1 - \frac{2 G M_G}{R_{\Omega} c^2}\right) ^{-1}
\end{equation}
The value $I_\Omega$ is significantly altered at high 
rotational rates. Hence, its derivatives with respect to
$\Omega$ and $M_G$ cannot be neglected in the equation for $\dot \Omega$.

The inner rim of the disk at $r_{in}$ is 
at   few NS gravitational radii, where relativistic gravity is
important. Therefore the specific angular momentum 
$l_{in}$ carried by matter differs from the classical newtonian value 
$\sqrt{G M_G r_{in}}$. Assuming an accretion disk co-rotating with the NS, 
and the Kerr metric for the approximate description of gravity around
a NS (see Shibata \& Sasaki 1998 for a more detailed
analysis of the metric), we obtain from Bardeen (1972):
\begin{equation}
l_{in} = c \frac{r_{h}^{1/2} (r_{in}^2-2 a r_{h}^{1/2} r_{in}^{1/2}
+ a^2)}{r_{in}^{3/4} (r_{in}^{3/2}-3 r_{h} r_{in}^{1/2} + 
2 a r_{h}^{1/2})^{1/2}}
\end{equation}
where  $r_{h} = G M_G/c^2$ and $a = I_{\Omega} \Omega / M_G c$. 

For an unmagnetized neutron star the inner rim $r_{in}$ of the accretion disk
either skims the NS surface or, for very compact NS, is truncated at the radius
of the last stable orbit $r_{ms}$.  
Therefore the radius $r_{in}$ is defined as:
\begin{equation}
r_{in} = \left\{ \begin{array}{cc} r_{ms} & \mbox{if $r_{ms} > R_{\Omega}$} \\ 
         R_{\Omega}  & \mbox{if $r_{ms} \le R_{\Omega}$}  \end{array} \right.
\end{equation}
\noindent
In the Kerr metric $r_{ms}$ is given by:
\begin{equation}
r_{ms} = r_{h} (3+Z_2-[(3-Z_1)(3+Z_1+2 Z_2)]^{1/2})
\end{equation}
with 
\begin{equation}
Z_1 = 1+ [1- (a/r_{h})^2]^{1/3} [(1+ a/r_{h})^{1/3}+(1- a/r_{h})^{1/3}]
\end{equation}
\begin{equation}
Z_2 = [3(a/r_{h})^2 + Z_1^2]^{1/2}.
\end{equation}

Solving for the differential equations (2) and (6), we determine the NS 
rotational evolution under the hypothesis of stationary accretion.
Our results are in agreement with the results of the relativistic code of 
CSTa, within $8\%$ even in the high spin regime (see {\it e.g.} Fig.~2). 
The related computational effort is greatly reduced 
with respect to a fully relativistic numerical approach.

\subsection{Magnetized case}

For the case of a magnetized NS, there is another characteristic length, 
i.e., the magnetospheric radius $r_{mag}$, which has to be compared with
$r_{ms}$ and $R_\Omega$ when calculating the value of $r_{in}$. 
The magnetospheric radius is computed as the product of the Alfven radius 
(see {\it e.g.} Lipunov 1992) times a factor $\phi \sim 1$, 
estimated as in Burderi {\it et al.} (1998). For high enough magnetic moment
$\mu$, $r_{mag}$ exceeds both $R_{\Omega}$ and $r_{ms}$ and the magnetic 
coupling between the disk and the star determines the extent of angular 
momentum transfer. The angular momentum balance relation in this case reads as 
$d(I_{\Omega} \Omega)/d t = g \dot{m}_B l_{in},$
where the torque function $g=g(\Omega)$ accounts for the details of
the interaction between the NS magnetosphere and the accretion disk.
When $g=0$ the NS is on the so--called ``spinup line'', where it can 
load mass without modifying its angular momentum $I_{\Omega} \Omega$. 
To bracket the uncertainties in the determination of the function $g$, we used
two different forms of it, following   Wang (1996) and Ghosh \& Lamb
(1991). In the integration scheme, equation (6) is modified accordingly,
to include the magnetic coupling. 

\section{The minimum mass load as a function of
$P$ for a magnetic NS }

We compute the $P$ vs $\ml$ relation for the case of an unmagnetized NS 
(as in CSTa) considering three selected EoSs. As illustrated in Fig.~3, 
the spin period $P$ shows a  steep dependence on the mass loading 
just above $\sim 3$ ms and small values of $\ml$ ($0.04-0.06 M_{\odot}$) are 
sufficient to produce a millisecond NS. By contrast, substantial mass load 
is requested for spinning the NS below the observed minimum period 
$\pmin = 1.558$ ms (PSR B1937+21). When approaching the limiting period 
$P_{lim}$ for mass shedding, the relation  $P~$vs$~\ml$ flattens. 
is $\sim 0.05 M_\odot.$  As already noted by CSTa, 
all EoSs (labeled as in Arnett \& Bowers 1977)
allow a NS to be spun up to $\pmin$, but a stiffer EoS requires a
higher value of $\ml$: $0.25~M_\odot$ for EoS L (stiff) as compared 
to $0.1~M_\odot$ for EoS A (soft). We have verified that such figures
are only slightly affected by assuming a lower initial gravitational mass
for the NS (e.g. 1.30 $M_\odot$ - 1.35 $M_\odot).$

In the case of a magnetized NS, the rate at which  angular momentum 
is transferred to the NS is controlled by the interaction of the magnetosphere
with the inner boundary of the accretion disk. Thus, the mass load 
$\ml$ which is necessary to attain a period $P$ depends upon
the magnitude of the $B$-field, on its evolution and on the torque 
function g($\Omega$). We describe different evolutionary 
pathways, during recycling, introducing two characteristic times: the
spinup time $\tau_{up} = P/{\dot{P}}$ and $\tau_{\mu} = \mu/{\dot{\mu}}$, 
the timescale of decay of the star's magnetic moment $\mu$. 

As a guideline we consider (i) a two-steps evolution in which a phase of 
significant $\mu$-decay (with $\tau_{\mu} \ll \tau_{up}$) precedes
the phase of spinup at constant $\mu$ (with $\tau_{\mu} \gg \tau_{up}$); 
and (ii) evolutionary pathways with $\tau_{\mu}=(6/7)~\tau_{up}.$
This condition describes the idealized situation of a NS sliding
parallel to the spinup line (corresponding to a fixed accretion
rate) without braking along it. In other words, the decay of $\mu$ is tuned 
just to allow for the NS to approach but never reach the equilibrium period, 
where the efficiency of the spinup process drops steeply (Wang 1996, 
Ghosh \& Lamb 1991). In fact, at the equilibrium period, the star would 
accrete matter without increasing its rotational rate. 
It is easy to demonstrate that the mass necessary to attain 
a final period $P$ in case (i) exceeds that of case (ii) by a factor
4/3. For $\tau_{\mu}\gg \tau_{up}$ even larger amounts of mass are needed.
Thus, case (ii) provides a {\it lower limit} on the mass load onto the NS, 
during the recycling process. 

The relativistic corrections at short rotational periods and the details of 
the disk-magnetosphere interaction prevent a simple analytical study of the 
spin evolution. Hence, we explored numerically the NS evolution. 
Fig.~4 collects a sample of pathways in the plane $P~$vs$~\ml$ obtained for 
the equation of state FPS (an intermediate EoS), combining various initial 
conditions (periods in the range $1-100$ sec, $\mu_{ini}$ between 
$10^{28} - 10^{30}$ G\,cm$^3$, accretion rates in the interval $0.01 - 1.00$ 
${\dot{M}}_{Edd}$) and selecting decay histories according to models
(i) and (ii). As illustrated in Fig.~4, the values requested to drive a NS 
to a final period $P$ depend on the magnetic field evolution.
We notice however, that only for NSs with $P>~5$ ms at the end of binary
evolution the mass load is sensitive to the history of the $\mu$-field.
For a magnetized NS the demand of accreted matter can at most halve, relative 
to the value inferred for a unmagnetized NS. If evolution proceeds to drive 
$P$ below $5$ ms, the request of $\ml$ depends weakly on the evolution of 
$\mu$ since the magnetospheric radius shrinks, becoming comparable to the 
radius of the last marginally stable orbit or to the physical radius of the 
star. Therefore, the $P~$vs$~\ml$ relation relaxes to the line 
that characterizes a unmagnetized NS.

\section{Discussion}

Millisecond pulsars (MSPs) with $P<~1$ ms have not been detected so far. 
Though such sources are beyond the sensitivity limits of the radio 
surveys conducted until now, available data may already place constraints 
on the distribution of MSPs at periods shorter than $\pmin$. 
Through  consideration of surveys sensitivities and known selection effects,
Cordes \& Chernoff (1997) carried out a likelyhood analysis and found
that the best-fit models are those increasing towards short periods 
with a best-fit minimum period that lies only slightly below $\pmin$.  
They in addition found a 95\% chance that the fastest MSPs are slower than 
1 ms, and a 1\% chance that they are as fast as 0.65 ms.
Recently, Possenti {\it et al.} (1998) carried out a population
synthesis calculation to determine the fraction of neutron stars with 
periods shorter than $\pmin$ relative to those having $P>\pmin$, for 
$\mu >7.3 \times 10^{25}$Gcm$^{3}$ (the minimum magnetic moment observed 
so far). They found that the process of recycling in low mass binaries can lead
to a distribution of periods extending below $\pmin$ (for the soft EoS 
a maximum is found to develop about 1 ms; Possenti {\it et al.} 1999, in
preparation). If future observations (as those undergoing at the Northern 
Cross in Medicina: D'Amico {\it et al.} 1998) will reveal such pulsars, 
the underlying NSs should have accreted at least $0.3 M_\odot$. 

Few estimates of the masses of MSPs in low mass binaries are now available 
and are reported in Fig.~5. The data, taken from Thorsett \& Chakrabarty 
(1998), are all compatible with the $P~$vs$~\ml$ relation derived in 
$\S~3$. In particular for PSR B1855+09, the MSP with a mass estimate 
based on GR alone ($M_G = 1.41 \pm 0.1 M_\odot$), the (minimum) mass requested 
to reach the observed period ($P=5.362$ ms) would be of at most 
$0.06 M_\odot$ (for the stiff EoS and an initially low value of $\mu$).

Thorsett \& Chakrabarty (1998) found that data
obtained collecting the mass measurements of all the known radio pulsars
in binary systems are consistent with a remarkably narrow underlying
gaussian mass distribution, with $M_{G}=1.35 \pm 0.04 M_\odot$. Their
sample of objects contains however five relativistic NS-NS binary systems
that contaminates the statistics of MSP binaries, narrowing the value of
$M$ and in turn of $\ml$. The progenitors of the NS-NS systems are massive
X-ray binaries, a population which does not lead to the formation of fast
spinning pulsars. If the proposed value of $\ml~<0.1~M_\odot$ is
representative of the MSPs observed sample, this figure is suggestive that
a fast decay of the magnetic field has occurred in these systems (not
only in PSR B1855+09). This is not in contraddiction with our estimates
on $\ml$ because only for those MSPs recycled to limiting periods $P<\pmin$ the
demand of mass load would exceed that of Thorsett \& Chakrabarty. 

Interestingly, a suggestion of the existence of NSs with masses
larger than $1.4 \msun$ arises also from a 
recent model (Stella \& Vietri 1999)
accounting for the pair kHz-QPOs observed in many LMXB 
sources (see {\it e.g.} van der Klis 1998 for a review). Stella \& Vietri
interprete the upper QPO frequency $\nu_2$ as due to matter inhomogeneities 
orbiting the NS at the inner disk boundary, while the lower QPO frequency 
$\nu_1$ is produced by the periastron precession at the inner edge of the 
accretion disk. In this framework, the observed clustering of the differences 
$\Delta \nu=\nu_2 - \nu_1$ around 250 - 360 Hz is naturally explained 
provided the gravitational mass of the typical accreting NSs is 
$\sim 2~M_\odot.$
Moreover the detection of millisecon X-ray pulsations from SAX J1808.4-3658
at different luminosity levels of the source suggests the presence of a 
massive NS in this system as discussed by Burderi and King (1998).

\section{Conclusion}

So far, there is no observational
evidence against the existence of submillisecond NSs. In the evolutionary
scheme adopted, the formation of submillisecond NSs is a natural outcome
if a minimum mass of $\simeq 0.25-0.30 M_\odot$ is accreted and if the
magnetic moment decays down to values $10^{25}-10^{26}$ Gcm$^3$ {\it
before} the end of recycling.

Only highly non-conservative scenarios may
indeed prevent the transfer of $\ml \ge 0.3~M_\odot$ in low-mass binaries.
In fact, evolutionary considerations shows that the mass transfer occurs
either because of the nuclear evolution or because angular momentum losses
of a $\sim 1~M_\odot$ companion (see {\it e.g.} Burderi, King \& Wynn
1996; Webbink, Rappaport \& Savonije 1983). 
A more massive companion ($M
\ga 2~M_\odot$ -- if we retain the hypothesis that the NS mass at birth is
$\sim 1.4~M_\odot$) would determine a dynamically unstable mass transfer
probably ending into a common envelope phase (see Kalogera \& Webbink
1996; 1998).  On the other hand, a low mass companion ($M \la 1~M_\odot$)
would either underfill its Roche lobe, because its nuclear evolution
timescale is longer than the age of the Galaxy for initial orbital periods
$\ga 0.5$ days, or, for shorter initial orbital periods, would fills its
Roche lobe during its main sequence lifetime, resulting in a
near--permanent faint LMXB whose compact object will never appear as a
radiopulsar. Therefore, in order to produce a fastly rotating NS, we are
restricted to a quite narrow range of companion masses, (namely
$1-2~M_\odot$), and of initial orbital periods (namely $0.5-10~$ days). 
For such binary systems, typical timescales for intense accretion are $\la
10^{8}$ years. Once the mass--transfer ceases, the endpoint is often a
circular binary containing a millisecond radio pulsar and a low--mass
($\sim 0.2-0.3 M_\odot$) white dwarf companion, the latter being the
degenerate helium core of the sub-giant which transferred mass to the
compact object (see {\it e.g.} Burderi, King \& Wynn 1998). This implies
that the companion has lost $\ga 1~M_\odot$ during the evolution of the
system. Even taking into account the case of an highly non-conservative 
mass transfer phase, it is difficult to believe that the NS in these systems 
has accreted less than $0.3~M_\odot$. 
                                                                               
There might be a number of physical processes that inhibit the
formation of sub-MSPs during binary evolution.

(a) Slow decay of the magnetic moment would relent the spinup process
thus requiring a amount of mass exceedingly large. In addition, precocious
freezing of $\mu$ (above $10^{27}$ G cm$^{3}$; as discussed in 
Konar \& Bhattacharya 1997; 1998) may prevent the formation of very fastly 
spinning NSs.

(b) Long term variation of the accretion rate may be important.
The terminal phases of LMXB evolution are still uncertain depending on
the Roche-Lobe feeding mechanisms.
If the accretion rate ceases on a time comparable to the evolutionary time 
scale, the propeller effect may cause an increase of the spin period above 
$\pmin$. Approximately, the ratio of the spin down time $\tau_{sd}$ by 
propeller to $\tau_{up}$ scales as
\begin{equation}
\frac{\tau_{sd}}{\tau_{up}}~=
~\left(\frac{{\dot{m}}_{up}}{{\dot{m}}_{sd}}\right)^{5/7}~
~\left(\frac{\mu_{up}}{\mu_{sd}}\right)^{2/7}~
~\frac{1}{\xi}
\end{equation}
where indexes $up$ and $sd$ refer to spinup and spindown, and $\xi$ is a 
parameter measuring the efficiency of the extraction of angular momentum 
from the NS during the propeller phase ($\xi=1.0$ if the efficiency of spin up
and down are equal). For spindown to be negligible, that ratio needs to exceed
unity, requiring a rather rapid turn off of the mass transfer rate
(${\dot{m}}_{sd} \ll {\dot{m}}_{up}$) (Possenti {\it et al.} in preparation). 

(c) Similar considerations holds in case in which the binary is subject to
disk instabilities (Cannizzo 1993 and references therein), leading to long 
periods of low level of activity -- referred in literature as quiescence -- 
interrupted by brief periods of outbursts (van Paradijs 1996, King 1998). 
For the known soft X-ray transients (SXTs: see Campana {\it et al.} 1998 for a
complete review) the first factor in equation (13) is $\sim 100$, whilst 
the ratio of $\mu$'s stays nearly constant during a cycle. 
If the quiescence phase lasts $\sim 100/\xi$ times longer than the
outburst phase, spin down would prevail. The off-on time ratio of observed
recurrent transients varies from source to source and is about  $10-100.$
It is thus difficult to assess the relevance of this process in affecting
the spin evolution and the statistical properties of the MSP population.
Recently Li, van den Heuvel \& Wang (1998) argued that unsteady mass 
transfer in PSR J1455-3330 may have occurred, and explored evolutionary tracks
allowing for spindown by propeller during the long quiescent intervals 
that the binary would experience if fed through a thermally unstable keplerian
disk. The discovery of SAX J1808-3658 might challenge this picture.

(d) For NSs with hot interiors, the viscosity of the nuclear matter may become 
inefficient in  damping the recently discovered $r-$mode instabilities
of the star (Andersson 1998). The related emission of gravitational waves 
(Stergioulas 1998) could limit the spin of a NS to a small fraction 
($\sim 0.2 - 0.3$) of its mass shedding rotational rate. For young NSs with 
$T > 10^9$, this effect could spindown the star at $P>10$ ms in a very short 
time ($\sim 1$ yr) (see {\it e.g.} Lindblom, Owen \& Morsink 1998). 
Andersson, Kokkotas \& Stergioulas (1998) claimed that a similar mechanism 
could occur during the LMXB-phase, thus preventing the spinup of a NS below 
$\sim \pmin$. However, the difficulties in the calculations of the values 
of the shear and bulk viscosity for superfluid nuclear matter at $T \sim
10^7-10^{8.5}$ K (relevant for the accretion phase) leave a large uncertainty
in the role of this instability in the recycling process. 
In addition,  crustal deformations induced by accretion
can excite emission of gravitational waves 
inhibiting the spin-up process to period shorter than $\sim 3$ ms  
(Bildsten 1998).

\newpage

\begin{table}[h]
\begin{center}
\vspace{0.5cm}
\begin{tabular}{c|c|c|c} \hline \hline
 EoS  &  $P_{\mathrm lim}$ (ms) &  $P_{\mathrm lim}$ (ms) & Difference \\ 
      &  (this work)            &(Cook {\it et al.}, 1994a)& (\%)    \\  \hline
  A   &   0.631                 &  0.604                  &  4.5     \\
  C   &   0.907                 &  0.894                  &  1.4      \\
  D   &   0.751                 &  0.730                  &  2.8      \\  
  E   &   0.677                 &  0.656                  &  3.1       \\    
  F   &   0.679                 &  0.715                  &  -5.1      \\  
  L   &   1.21                  &  1.25                   &  -2.9       \\
  M   &   1.48                  &  1.49                   &  -0.9       \\  
  N*  &   1.03                  &  1.08                   &  -4.7       \\    
  KC  &   0.910                 &  0.888                  &   2.5       \\  
  AU  &   0.708                 &  0.701                  &   1.0      \\
  UU  &   0.792                 &  0.784                  &   1.0      \\
  UT  &   0.764                 &  0.754                  &   1.4      \\
  FPS &   0.767                 &  0.747                  &   2.8      \\
\hline \hline
\end{tabular}
\caption{Minimum rotational periods for the EoSs indicated in column~1.
In column~2 we give the periods obtained from the analytical model derived in 
this work. In column~3 we collect the numerical values from Cook, Shapiro, 
\& Teukolsky (1994a). The fourth column indicates the percentage differences.}
\label{periods}
\end{center}
\end{table} 

\newpage


\begin{figure}
\centering
\psfig{file=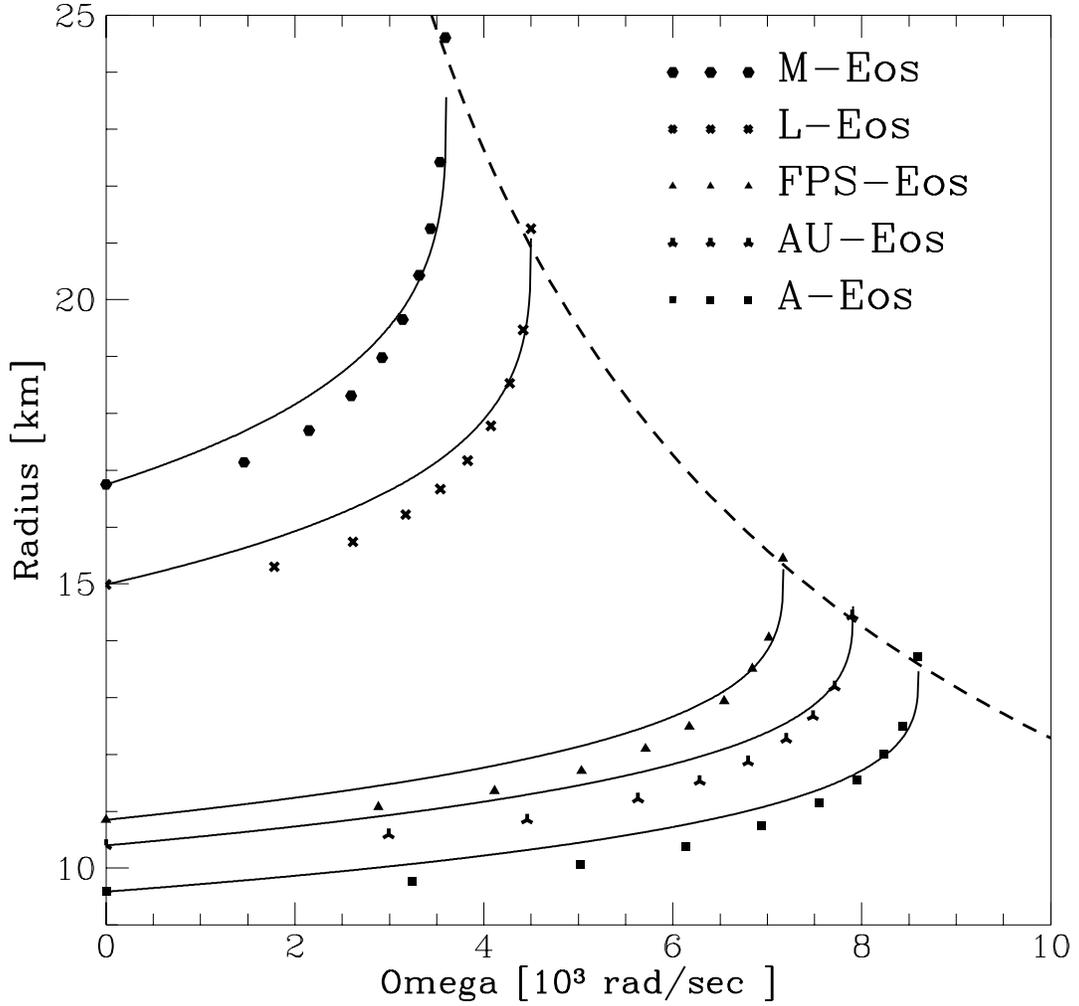,width=15cm}
\caption {
NS circumferential radii (in units of km) versus
angular frequency $\Omega$ (in units of $10^3~$ rad/sec). The points are from 
Cook, Shapiro \& Teukolsky (1994b) for 5 representative EoSs
$(M_{B}~=~1.4~M_\odot).$ The {\it solid lines} are relation (5). The 
{\it dashed line} is the keplerian relation (3).
}
\end{figure}

\begin{figure}
\centering
\psfig{file=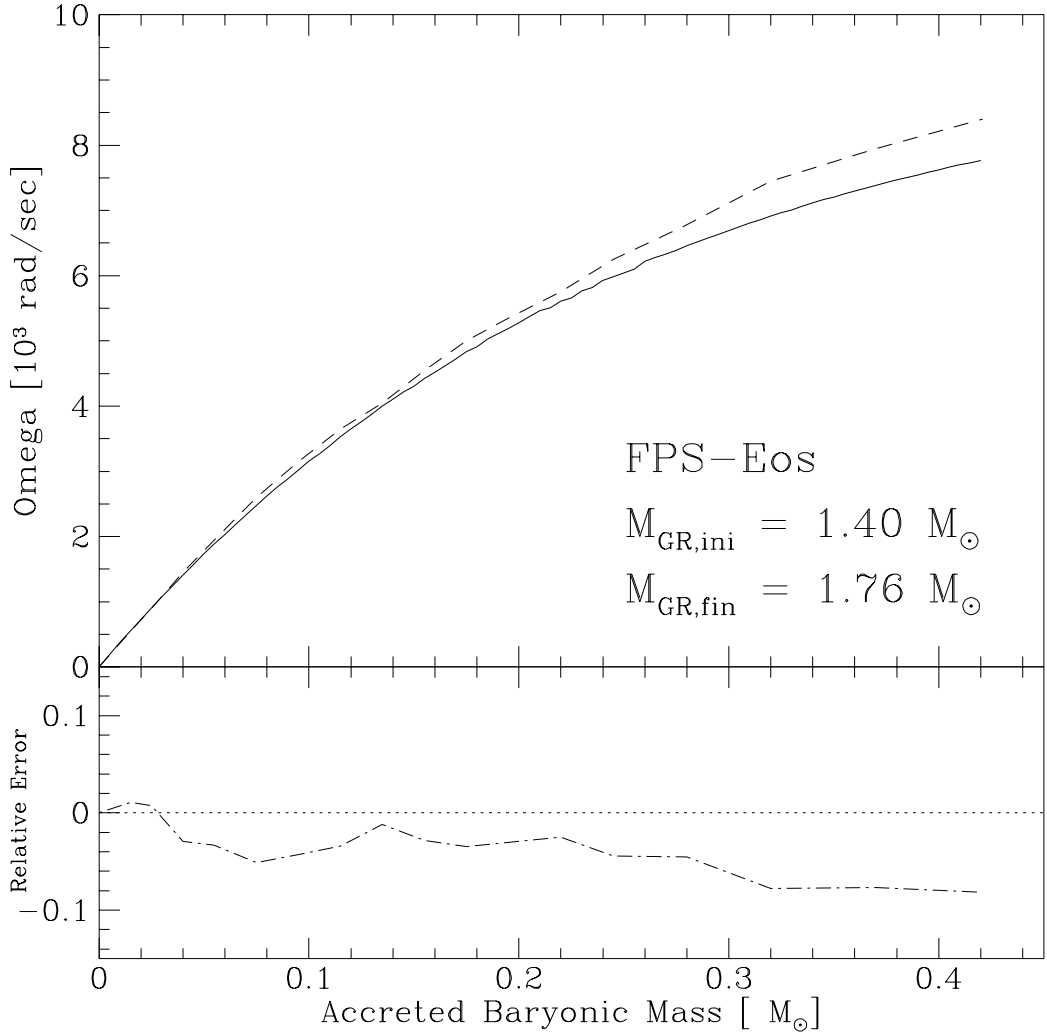,width=15cm}
\caption {
Rotational evolution of an unmagnetized NS (initial 
gravitational mass $M_{G,ini}~=~1.40~M_\odot$) undergoing steady 
accretion of matter. The $\Omega~$vs$~M_{ac}$ relation calculated for FPS-EoS
with the semi--analytic model described in the text ({\it solid line}) is 
compared to that from CSTa ({\it dashed line}). The relative errors in 
computing $\Omega$ with our approximate model are displayed in the lower 
panel. Accretion is halted at $M_{ac}~=0.416~M_\odot,$ when mass shedding sets 
in (the final gravitational mass of the NS is $M_{G,fin}~=~1.76~M_\odot).$ 
}
\end{figure}

\begin{figure}
\centering
\psfig{file=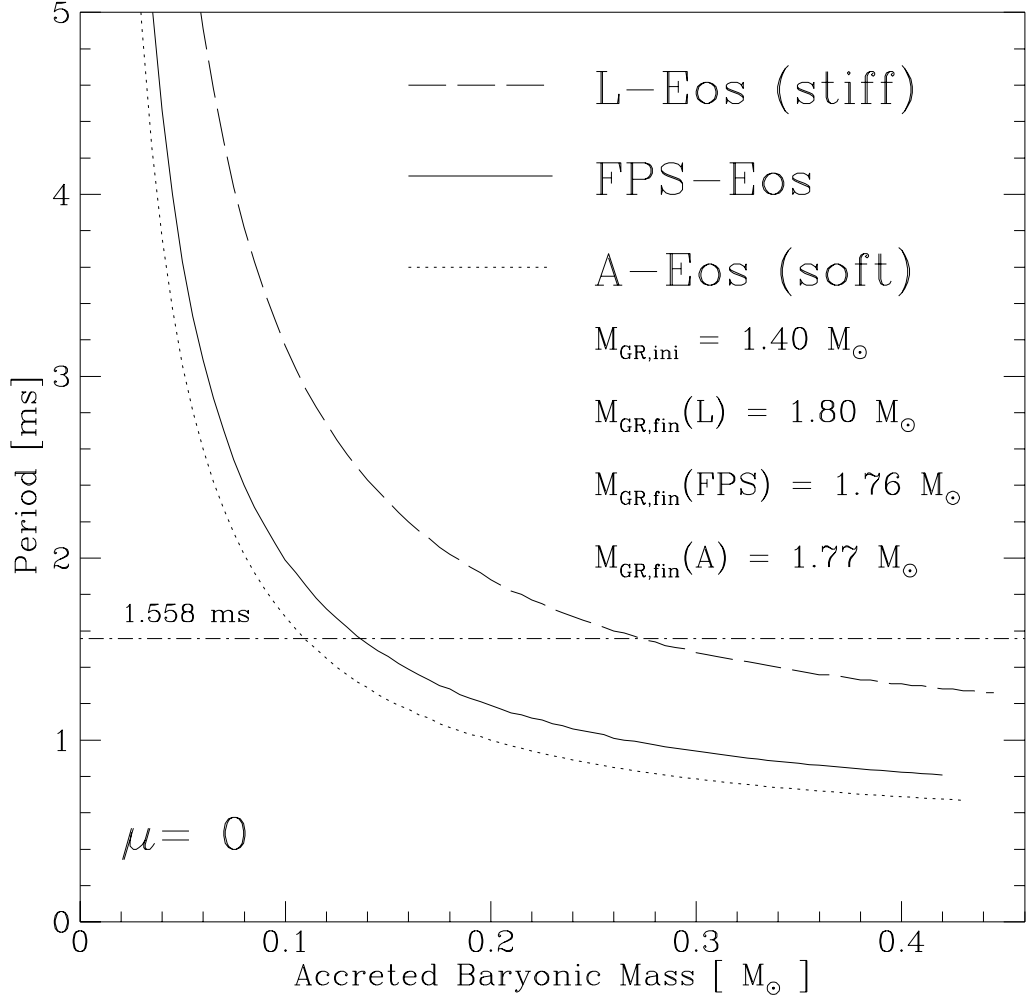,width=15cm}
\caption { $P~$vs$~\ml$ relations for the case of a unmagnetized NS 
considering three selected equations of state: L (stiff: {\it dashed line}), 
A (soft: {\it dotted line}), FPS (intermediate: {\it solid line}). The initial
gravitational mass of the NS is always $1.40~M_\odot.$ The period $~P$ is 
measured in millisecond, the accreted baryonic mass $\ml$ is in solar mass. 
The dashed--dotted line represents the minimum rotational period observed so 
far. Accretion is halted when the centrifugal limit is reached.
}
\end{figure}

\begin{figure}
\centering
\psfig{file=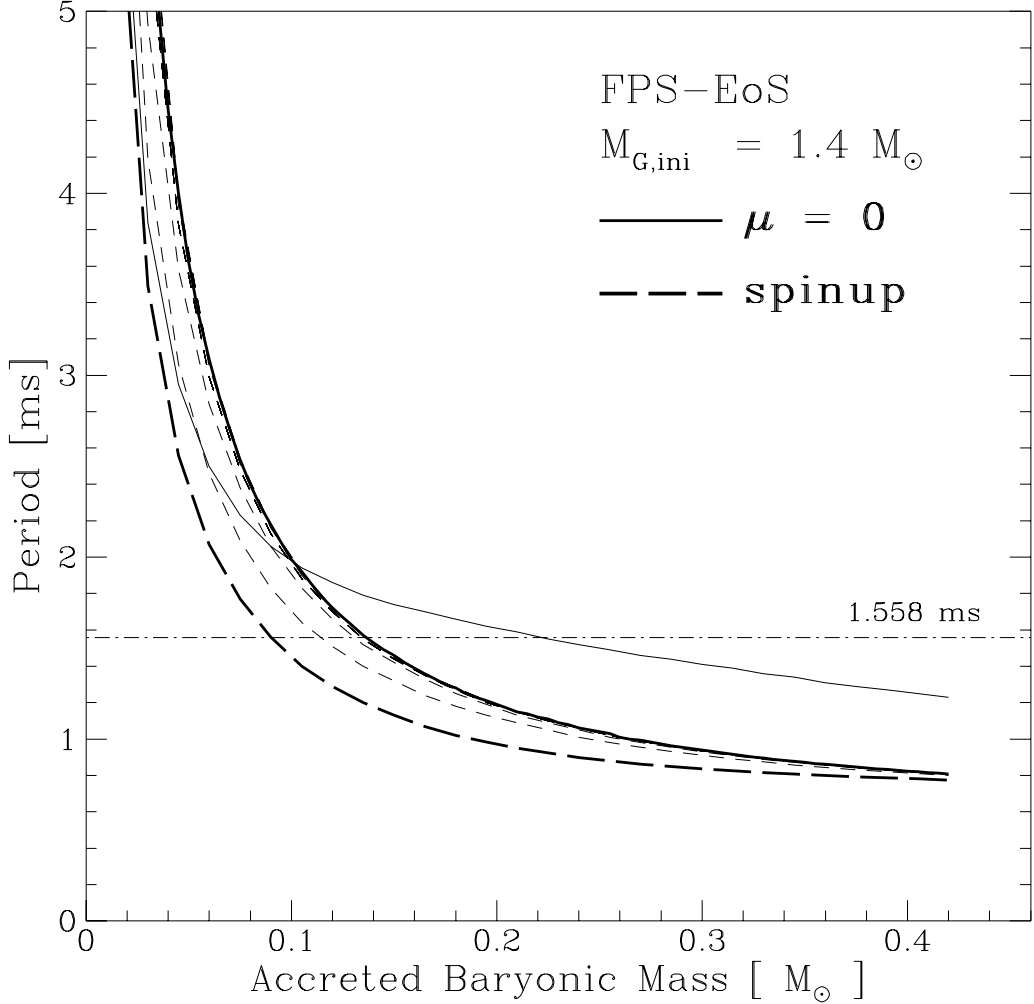,width=15cm}
\caption {
$P~$vs$~\ml$ relations for a magnetized NS using FPS EoS and 
an initial gravitational mass of $1.40~M_\odot$. The different pathways 
(a sample is represented by the {\it dashed lines}) define a strip, which 
narrows towards shorter periods (see text for details). The strip is upper 
bounded by the evolutionary path for an unmagnetized NS ({\it bold solid 
line}). The {\it bold long dashed line} refers to the evolution along the
spinup line and is calculated assuming a step shaped $g$-torque function 
(an extreme situation which maximizes the efficiency of the 
spinup process). The {\it thin solid line} represents the effect of a slow
$\mu$-decay: the evolution along the spinup line is ``braked'' and the
requested mass load increases.
}
\end{figure}

\begin{figure}
\centering
\psfig{file=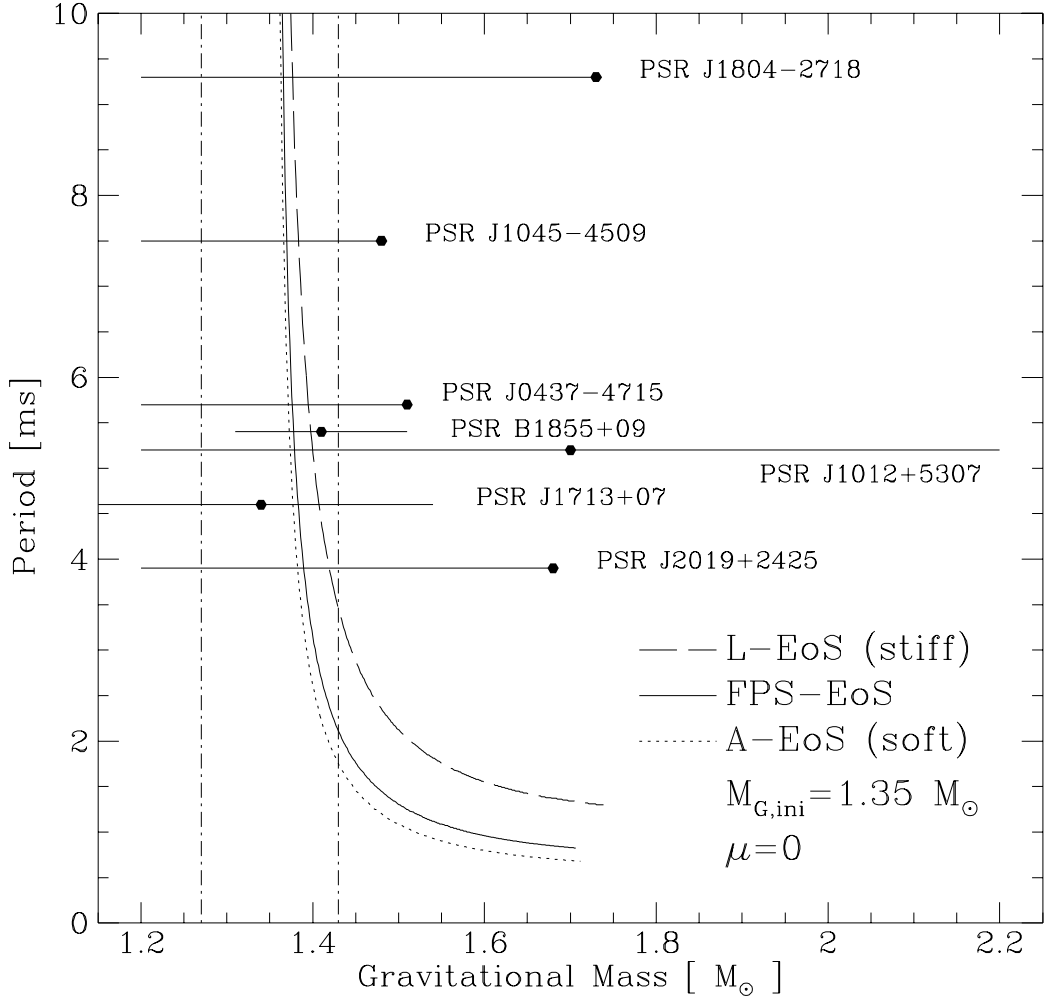,width=15cm}
\caption {
Available mass estimates for MSPs ($P<10$ ms)
compared with the evolutionary lines labeled as in Fig.~3, but calculated for
an initial gravitational mass of $M_{G,ini}~=~1.35~M_\odot.$
The error bars indicate central $68$\% confidence limits, while 
upper limits are one-sided $95$\% confidence limits. All the estimates are 
from Thorsett \& Chakrabarty (1998). Vertical lines bind the 
$2\sigma$-confidence region for the NSs masses ($M_{G}~=~1.35 \pm 0.08$) 
derived by the same authors.
}
\end{figure}
	
\end{document}